\documentclass[aps,prb,showpacs,twocolumn]{revtex4-1}
\usepackage[pdftex,plainpages=false,colorlinks=true,linkcolor=blue, citecolor=blue, urlcolor=blue]{hyperref}

\usepackage{amssymb}
\usepackage{epsfig}
\usepackage{graphicx}
\usepackage{amsmath}
\usepackage{array,color}
\usepackage{natbib}

\begin{document}

\title{ Phase diagram and Fermi-liquid properties of the extended Hubbard model on the honeycomb lattice}

\author{Wei Wu$^{1}$ and A.-M.-S. Tremblay$^{1,2}$}
\affiliation{
$^1$D{\'e}partement de Physique and RQMP, Universit{\'e} de Sherbrooke, Sherbrooke, Qu{\'e}bec, Canada \\
$^2$Canadian Institute for Advanced Research, Toronto, Ontario, Canada
}

\date{\today}

\begin{abstract}
The Hubbard model and extended Hubbard model on the  honeycomb lattice can be seen as prototype models of single layer graphene placed in a high dielectric constant  
environment that screens the Coulomb interaction.
Taking advantage of the absence of a sign problem at half-filling, we study this problem with clusters up to 96 sites with the Determinant Quantum Monte Carlo Method as an impurity solver for the the Dynamical Cluster Approximation at finite temperatures. After determining the stability of the semi-metallic phase to interaction-induced spin-density wave (SDW), charge-density wave (CDW) and Mott insulating phases, we study the single particle dynamics of the Dirac fermions. We show that when spontaneous symmetry breaking is avoided, the semi-metallic phase is a stable Fermi liquid in the presence of repulsive interactions and that Kondo screening dominates
the low temperature regime, even though there is a $\rho(\omega) = |\omega|$ type local density of states.
We also investigate the impact of the correlation effects on the renormalization of the Fermi velocity $v_F$. We find that $v_F$ is not renormalized when only on-site repulsion $U$ is present, but that near-neighbor repulsion $V$ does renormalize $v_F$. This may explain the variations between different measurements of $v_F$ in graphene.   
\end{abstract}


\maketitle

\section{introduction}
Graphene, fabricated as a novel two dimensional material, is considered as a promising material for future electronics. 
In graphene physics, the Fermi velocity $v_F$ is an important quantity that determines various fundamental physical properties of this system.~\cite{neto2009electronic} Generally speaking, the renormalization of $v_F$ from its bare 
value reflects the effective interaction strength in graphene.

In recent years, great progress has been achieved to experimentally measure $v_{F}$ in graphene. There are still puzzling questions however about the measured values of $v_{F}$ placed on different substrates.
For graphene in vacuum, namely suspended graphene, D. C. Elias \textit{et al}~\cite{elias2011dirac} found that $v_{F}$ increases logarithmically 
as one approaches the Fermi level, eventually reaching $v_{F}  \approx 3.0  \times 10^{6}ms^{-1} $, more than triple the bare value  near the charge neutrality point. 
This finding apparently confirms the occurrence of strong correlation effects in this system. This validates random phase 
approximation calculations~\cite{sarma2007many} based on
the usual assumption that due to the poor dielectric screening, Coulomb interaction between electrons are 
long range and rather strong.~\cite{wehling2011strength} Similiar logarithmically renormalized $v_{F}$ is also observed in
gaphene on hexagonal boron nitride (hBN) surface~\cite{chae2012renormalization}
though the largest $v_F$ detected there is only about $1.3 \times 10^{6}ms^{-1}$.
Nevertheless, there are many other experiments that actually endorse weak or barely enhanced
$v_{F}$ in graphene, regardless of the different substrate dielectric constants. ~\cite{martin2007observation, yankowitz2012emergence, reed2010effective}
In particular, by using angle-resolved photoemission spectroscopy (ARPES) method, a
nearly ideal linear band structure was observed for isolated multilayers of graphene on SiC substrate \cite{sprinkle2009first},
with a $v_{F}  = 1.0 \pm 0.05 \times 10^{6}ms^{-1} $. This implies that correlation effects are irrelevant for the single particle 
dynamics in this system.

In this paper, we study this problem using the Hubbard and extended Hubbard model 
on the honeycomb lattice. One first needs to find the region of stability of the semi-metallic phase in parameter space, so we first determine the phase diagram. Most recently, both analytical and numerical studies of the phase diagram mainly focused on the
Mott transition and possible emergent exotic quantum states.~\cite{wu2010interacting, liebsch2013coulomb, meng2010quantum, assaad2013, sorella2012absence} In particular, much attention has turned to
the controversial topic of the existence of a $Z_2$ spin liquid phase laying between 
the semi-metal and the antiferromagnetic state.~\cite{meng2010quantum, assaad2013, sorella2012absence}.
Here we focus on the different phase transitions and single particle dynamics of the interacting Dirac fermions 
that are relevant to the low-energy physics of single-layer graphene. We find in passing that the $Z_2$ spin-liquid is pre-empted by an antiferromagnetic quantum critical point.
 
Taking advantage of the absence of a sign problem at half-filling, we attack these problems by solving a cluster extension of dynamical mean field theory for clusters up to 96 sites 
using mostly a large scale determinant quantum Monte Carlo (DQMC) method as an impurity solver. 

The model and method are described in the following section. The main results are presented in three subsections of Sec.~\ref{Sec:Results}: First we find  phase diagrams as a function of temperature and interaction strength, second we show that in the semi-metallic regime, the low temperature behavior is that of a Fermi liquid, and finally, we study the effects of interactions on the Fermi velocity $v_F$.  The results are summarized and briefly discussed in Sec.~\ref{Sec:Summary}.

\section{model and method}
\begin{figure}[!t]
\begin{center}
\includegraphics[width=8cm]{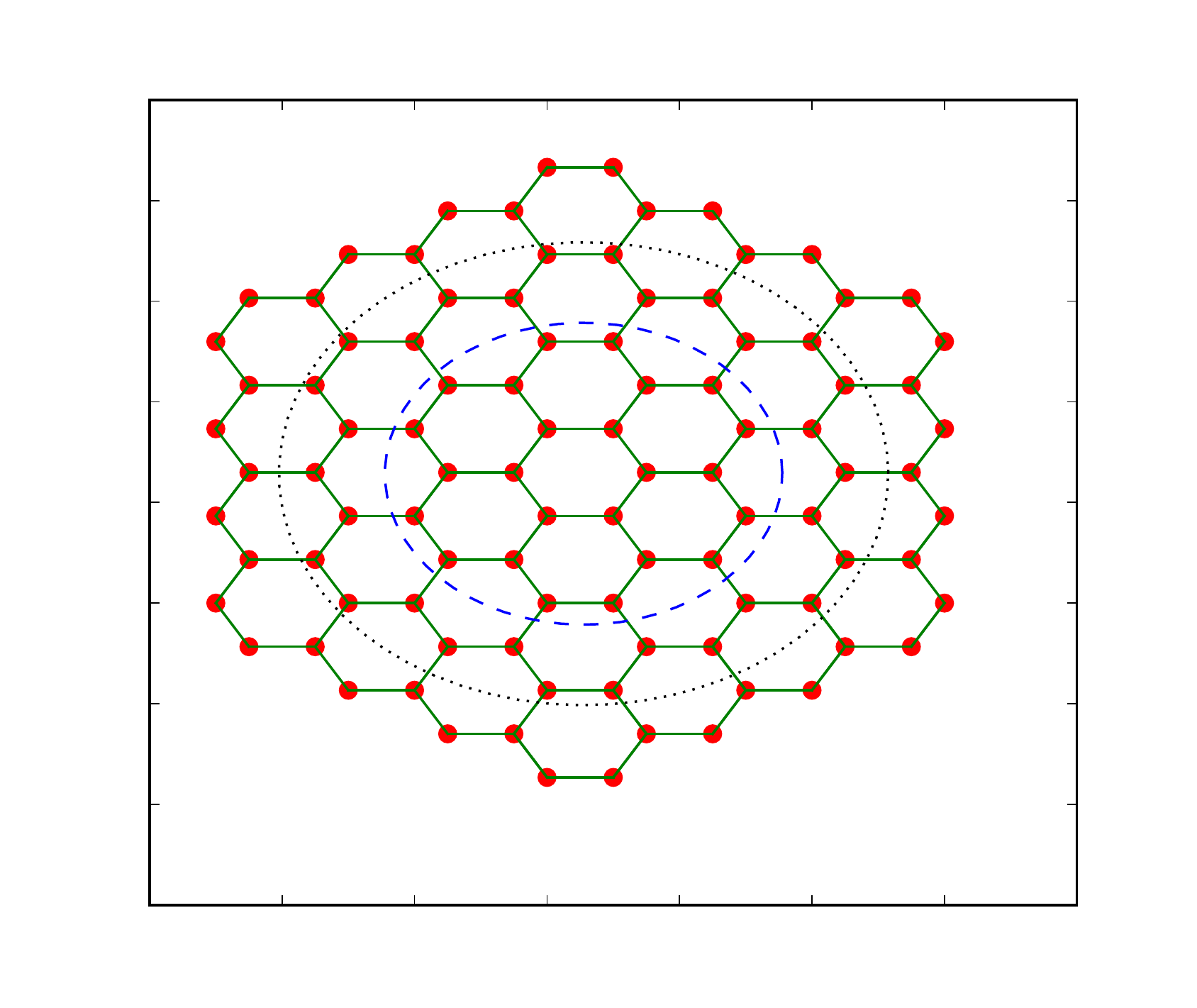}
\end{center}
\caption{(Color online) Illustration of the 96-site cluster studied in this work. 
Dashed (dotted) line encircles the 24-site (54-site) also used in our calculations. 
}
\label{fig:cluster}
\end{figure}

We investigate the extended Hubbard model on the honeycomb lattice defined by
the Hamiltonian,
\begin{eqnarray*}
H = -t \sum_{<i,j>,\sigma} c^{\dagger}_{i\sigma} c^{\null}_{j\sigma}
+ U \sum_{i} n_{i \uparrow}n_{i \downarrow} \\
+ \sum_{i,j,\sigma , \bar{\sigma} } V_{ij}  n_{i\sigma}n_{j \bar{\sigma} } 
-\mu \sum_{i} c^{\dagger}_{i\sigma} c^{\null}_{i\sigma},
\end{eqnarray*}
where $c^{\dagger}_{i\sigma}$ and
$c_{i\sigma}$ are creation and annihilation operator for fermions with site index $i$
and spin index $\sigma$, $n_{i\sigma}=c^{\dagger}_{i\sigma} c_{i\sigma}$ is the density operator, $t$ is hopping amplitude 
between nearest-neighbor sites $<i,j>$, while $U$ is the on-site repulsion and $V_{ij}$
is the Coulomb repulsion between occupied sites. We assume $i > j$ to eliminate double counting of non-local repulsion $V$.
The standard Hubbard model corresponds to $V_{ij}=0$. In our study, due to the 
limitations of the determinant quantum Monte Carlo (DQMC) \cite{blankenbecler1981monte,loh1992stable} impurity solver, we only cope with nearest-neighbor (NN) and the next-nearest-neighbor
(NNN) repulsions. The chemical potential $\mu$ is always chosen as $-\frac{U}{2}- \sum_{j}\frac{V_{ij} }{2}$ to fulfill the half-filling condition. We work in units where hopping $t$ is unity.

The tight-binding limit ($U,V_{ij}=0$) of this model contains the famous massless Dirac fermions at low energies which are extensively 
employed as a first approximation for studies of single layer graphene.\cite{wallace1947band, neto2009electronic}

Here, we mainly employ the dynamical cluster approximation (DCA) method~\cite{Hettler:1998} combined with large-scale DQMC impurity solver~\cite{KhatamiSolver:2010} throughout
our study. The prominent merit of DQMC as an impurity solver is that the quasi-linear scaling of computing time with respect to the inverse
temperature~\cite{KhatamiSolver:2010} $\beta$ allows one to reach large system sizes at low temperature. 

In our study, a DCA effective impurity model consisting of 96 cluster sites (see Fig. \ref{fig:cluster}) and about 200 bath sites is simulated at temperatures as low as $T \sim 1/30$. By using a self-adaptive cluster-bath coupling scheme we are able to keep the parametrization error of the Weiss functions of DCA
below $0.1\%$, which is negligible compared to the typical statistical error of about $1\%$ in DQMC simulations. This statistical error results from, typically, 
$2\times 10^{5}$ to $5 \times 10^{5} $ Monte Carlo sweeps for each DCA iteration. 
In addition to the DQMC, the weak coupling expansion of continuous time quantum Monte Carlo (CTQMC) \cite{rubtsov2005continuous} is also used as impurity solver for  calculations on smaller clusters 
but much lower temperatures to complement our study.

\section{Results and analysis}\label{Sec:Results}
We first present our results for the phase diagram and then discuss in the remaining two subsections the Fermi liquid regime and the influence of interactions on Fermi velocity renormalization. 
\subsection{Phase transitions}
\begin{figure}[!t]
\begin{center}
\includegraphics[width=9cm]{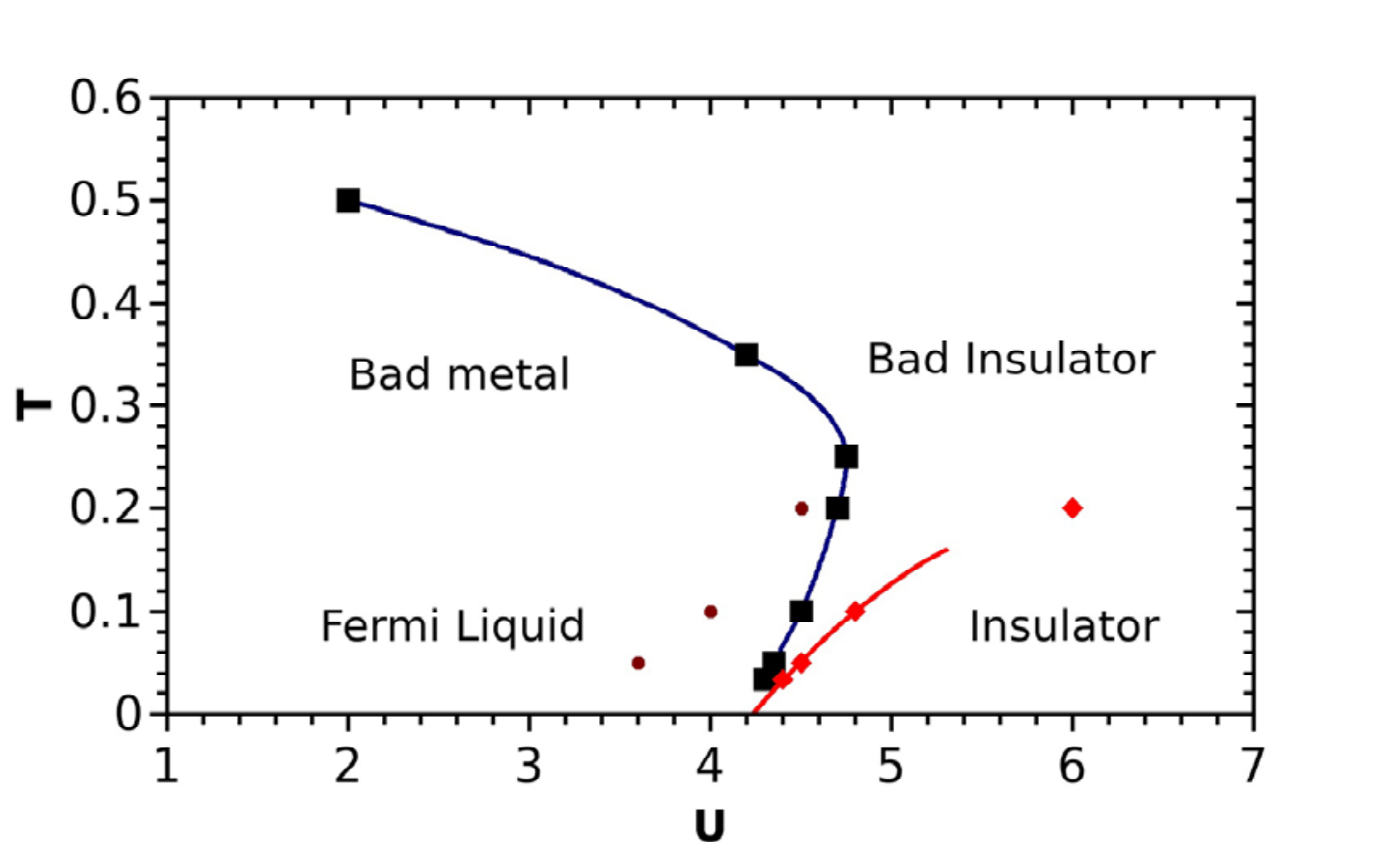}
\end{center}
\caption{(Color online) Finite temperature phase diagram of the Hubbard model ($V_{i,j}=0$) on the honeycomb lattice obtained with a 96-site DCA calculations.
The Mott transition from semi-metal to insulator at zero temperature is estimated to be $U_{Mott} \thickapprox 4.2$ by extrapolating the red diamonds. 
Brown circles denote the semi-metal to antiferromagnetic spin-density wave (SDW) phase transition found in the same calculation but without the restriction to a paramagnetic state. We estimate the position of the zero-temperature phase transition as $U_{c}^{SDW} \lesssim  3.6t$. The blue squares indicate the point where the imaginary part of the self-energy becomes flat at small Matsubara frequencies, distinguishing the bad metal and bad insulator phases \cite{park2008cluster}. Lines are guides the eye.
}
\label{fig:phaseDiagram}
\end{figure}

The finite temperature phase diagram for the standard Hubbard model on the honeycomb lattice, \textit{i.e.}, when $V_{ij} = 0$ and $U \neq 0$, is shown in Fig. \ref{fig:phaseDiagram}. 
It was obtained with a 96-site DCA calculation. 
The transition point to the paramagnetic Mott insulator at zero temperature is estimated to be $U_{Mott}\thickapprox 4.2t$ when paramagnetic conditions
are artificially imposed. If we lift this restriction and allow for a spin density wave (SDW) transition, we find that it appears at $U_{c}^{SDW} \lesssim  3.6t$. A summary of estimates for the zero-temperature phase transitions found with other methods can be found in Refs.~\onlinecite{liebsch2013coulomb,Arya:2014}. 

For the SDW phase transition our value $U_{c}^{SDW} \lesssim  3.6t$ is a bit smaller than the large scale quantum Monte Carlo results  $U_{c}^{SDW}\thickapprox3.8$ in Ref.~\onlinecite{assaad2013} and 
  $U_{c}^{SDW}\thickapprox3.87$ in Ref.~\onlinecite{sorella2012absence}. The differences in the $U_{c}^{SDW}$
is a direct consequence of the mean-field nature of dynamical mean-field theory methods. We expect that $U_{c}^{SDW}$ for the semi-metal to antiferromagnetic transition would systematically approach the large scale QMC value upon increasing the cluster size of DCA. 
 
In Ref.~\onlinecite{assaad2013} it was found with large scale quantum Monte Carlo calculations that $U_{Mott}\thickapprox 3.4$. This suggested a possible regime between $U_{Mott}\thickapprox 3.4$ and $U_{c}^{SDW}$ where a spin-liquid phase could exist. In our calculation, the spin-liquid phase is pre-empted by the appearance of antiferromagnetism since our estimate for the critical value $U_{Mott}\thickapprox 4.2t$ is larger than that found for the antiferromagnetic phase $U_{c}^{SDW} \lesssim  3.6t$. The non-existence of a zero-temperature spin-liquid phase for this model is in agreement with the results of larger scale Quantum Monte Carlo calculations\cite{sorella2012absence} and with other cluster-in-a-bath calculations.~\cite{hassan2013} Convergence with cluster size is, however, subtle for this problem.~\cite{liebsch2013coulomb,WuLeHur:2012}

  Since screening in graphene is not expected to be very effective because of the low density of states at the Fermi level, it is useful to consider the phase diagram in the presence of both on-site interaction $U$ and nearest-neighbor inter-site interactions $V$. This
reveals a competition between the antiferromagnetic SDW phase and the staggered charge density wave (CDW) phase
that has non-equivalent electron density on the A/B sublattices and hence also breaks chiral symmetry, like the antiferromagnetic phase.

\begin{figure}[!t]
\begin{center}
\includegraphics[width=10cm]{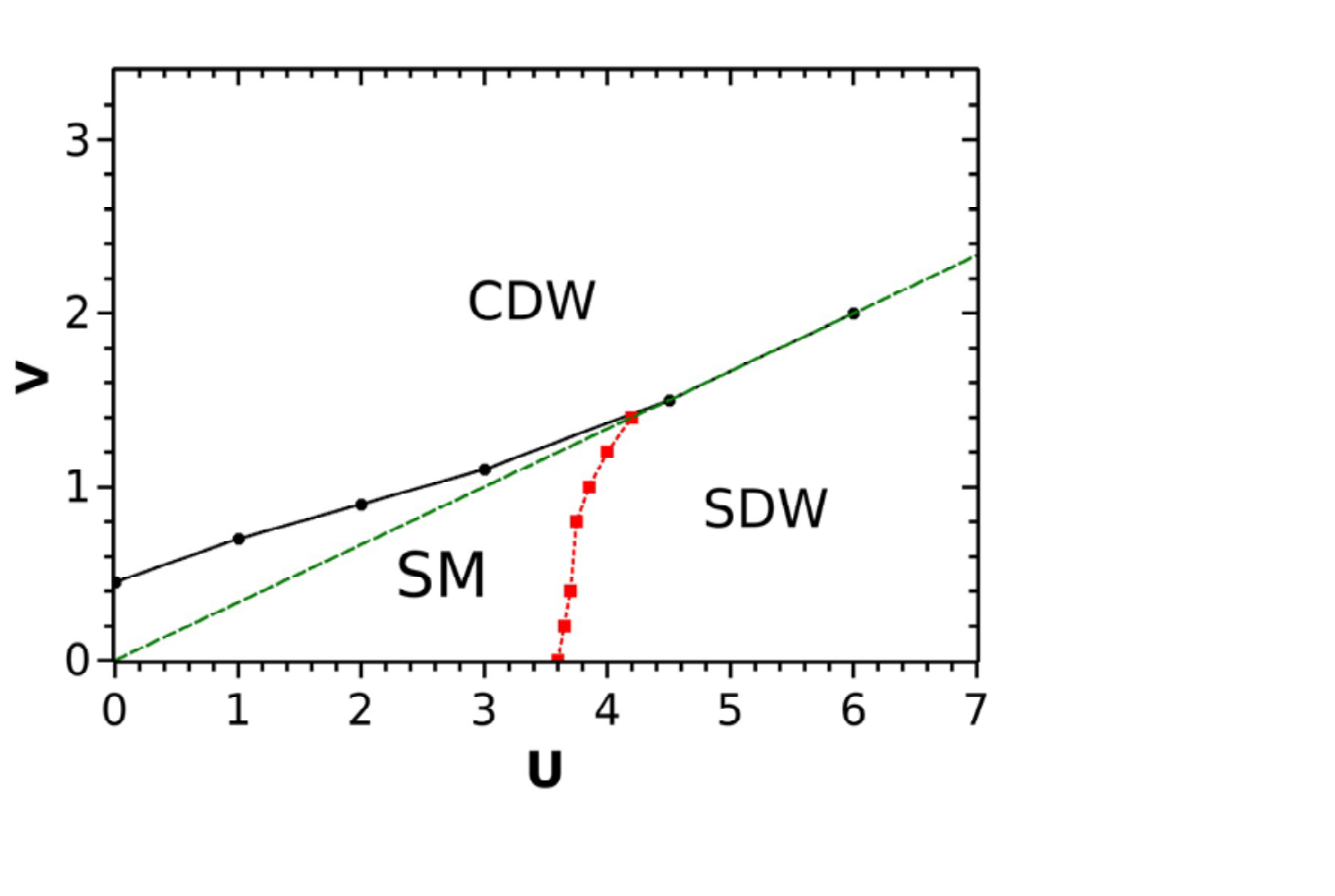}
\end{center}
\caption{(Color online) Phase diagram of the extended Hubbard model on the honeycomb lattice. $V$ is the nearest-neighbor repulsion. Data points are obtained within a 24-site DCA calculation at $T=0.05$.
The correlation length needed to establish long-range order on a smaller cluster leads to a slightly underestimated $U_c \sim 3.6$ for the SM/SDW transition at $V=0$. 
The locus $U=3V$ is indicated with a dashed green line. The phase transition between semi-metal (SM) and antiferromagnetism (SDW) is continuous (dotted red line with red square symbols) while SM/CDW and
SDW/CDW transitions (solid line black line with black circles) are of first order. The staggered
CDW phase is defined as a state where electron densities become unequal on the A/B sublattices.
}
\label{fig:UV}
\end{figure}

The phase diagram is displayed in Fig.~\ref{fig:UV}. When $U, V \gg t$, the physics is controlled mostly by the competition between $U$ and $V$. In that limit, 
as expected from simple classical arguments on potential energy, a phase transition between SDW and CDW occurs around $U \sim zV$, where $z$ is the coordination number
of the lattice, namely $z=3$ here. Consistent with this result, one has $U_c \approx 4V$ on the 2d square lattice \cite{zhang1989extended} and $U_c \approx 2V$ on the 1d chain \cite{hirsch1984charge}. However, for small
$U$ and $V$, the hopping term $t$ takes part in the competition, resulting in a CDW to semi-metal (SM) transition at $U<zV$. The phase transition ends at $U=0$,
$V=0.45$, where the charge density fluctuations solely drive a CDW transition. 

It is notable that the phase transitions involving the
CDW, namely the SDW/CDW and SM/CDW phase transitions, are all of first-order while the SM/SDW transition is continuous. This
behavior is reminiscent of the analogous phase transitions on the 2d square lattice \cite{aichhorn2004charge} but is different from the 1d chain case \cite{Fourcade:1984,hirsch1984charge, aichhorn2004charge} 
where the transition eventually becomes continuous as $U,V$ decrease. Note that in mean-field theory~\cite{raghu_topological_2008} the SM/CDW transition was found to be continuous, in contrast with our more reliable calculation. 

 \subsection{Fermi liquid}
  While the critical point for the Mott transition in Fig. \ref{fig:phaseDiagram} control a large area of 
  the bad insulator and bad metal region at finite temperatures, it actually does not engulf the whole small $U$ 
  regions of the phase diagram. In other words, a semi-metallic state persists at low temperatures against the formation 
  of local moments when $U$ is sufficient small that the Mott and the SDW transitions are avoided. 
  
  Is this semi-metallic state a Fermi liquid? It has been argued that owing to the 
  linearly varnishing low energy density of states, the Hubbard model on the honeycomb lattice may possess an under-screened fixed point \cite{bulla1997anderson}. 
  This argument seems to agree with numerical renormalization group analysis, \cite{chen1995kondo,gonzalez1998renormalization} which shows that when an Anderson 
  impurity site couples to a pseudogap conduction band with density of states $\rho\sim|\omega|^{r}$, the local moments on the impurity sites
 cannot be fully screened  out even at zero temperature when $r>\frac{1}{2}$, because the itinerant electrons are exhausted. Based on
 this argument, a non-Fermi liquid
 should be expected on the honeycomb lattice, since at low-energies it has $r$ equals $1$. 

\begin{figure}[!t]
\begin{center}
\includegraphics[width=9cm]{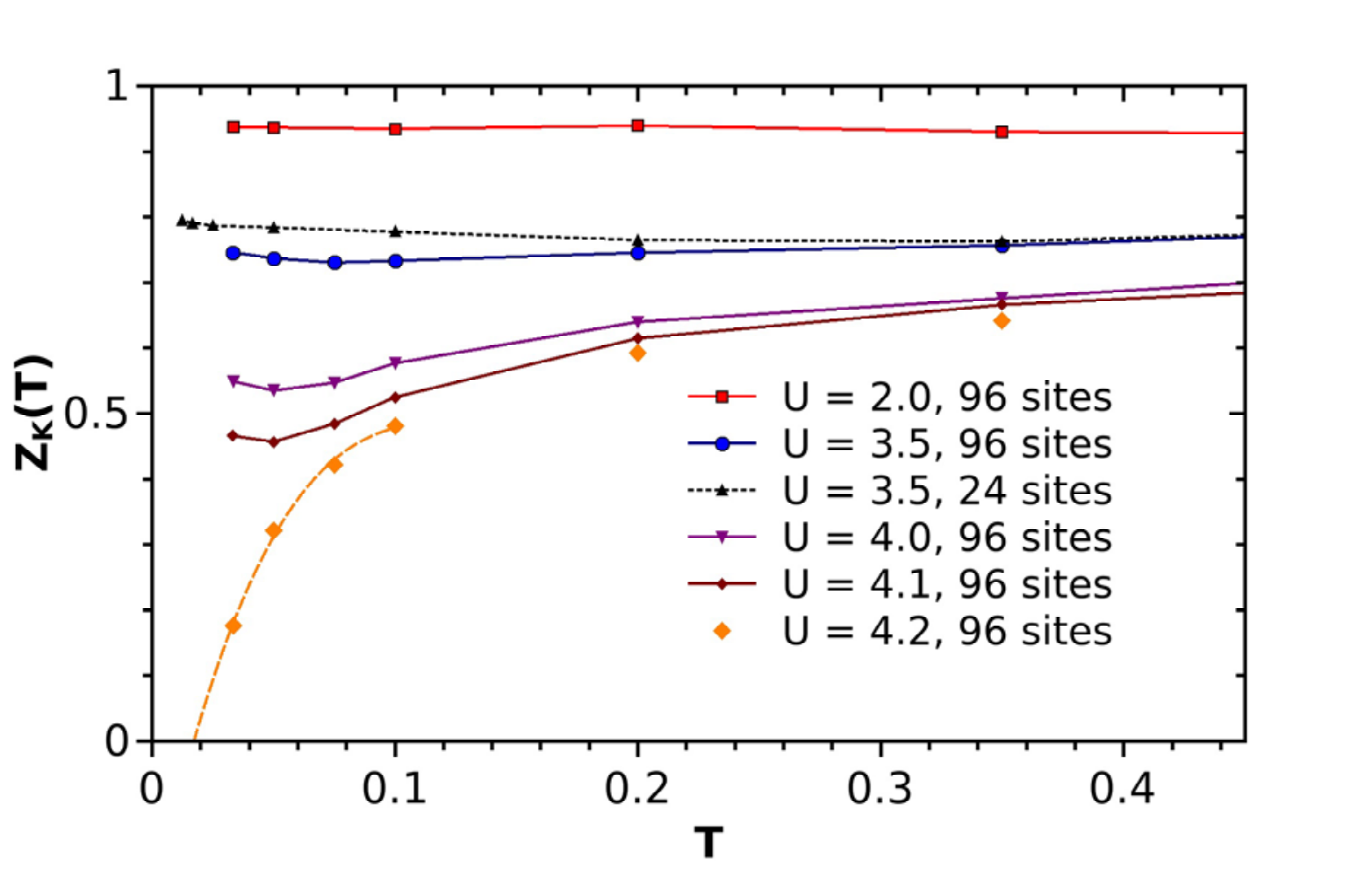}
\end{center}
\caption{(Color online)  $Z_{T}(k)$ as a function of temperature $T$ for different $U$ of the Hubbard model. Polynomial fitting of the $U_c = 4.2$ line suggests where the system becomes insulator. For $U<4.2$, the quasi-particle residue is finite hence suggesting a Fermi liquid state. A 24-site DCA + CTQMC calculation was also used at $U=3.5$ to reach the very low-temperature properties of $Z_{T}(k)$.
}
\label{fig:residue}
\end{figure} 
 However, our DCA simulations unambiguously 
  suggest the existence of a Fermi liquid at low temperatures. 
  In order to clarify this paradox for the single particle behavior of the Dirac fermions, we investigate the following physical quantity~\cite{vidhyadhiraja2009quantum},
\begin{equation}
 Z_{T}(k)=\frac{1}{1-\frac{Im\Sigma(k,\pi T)}{\pi T}}.
 \end{equation}
 which becomes the quasiparticle renormalization factor in the limit $T\rightarrow0$. For a Fermi liquid phase, 
this quantity stays finite at zero temperature, 
in contrast with the marginal Fermi liquid \cite{vidhyadhiraja2009quantum} or Mott insulator where $Z_{T}(k)$ goes to zero in the limit $T\rightarrow0$. 
Fig. \ref{fig:residue} shows $Z_{T}(k)$ as a function of $T$ in the semi-metallic phase (no SDW allowed) for various values of $U$ at the Dirac point $K$. From this, we can see that for $U<4$
  the quantity $Z_{T}(k)$ varies slowly as the temperature decreases reflecting the Fermi liquid characteristic. By further increasing $U$, $Z_{T}(k)$
  becomes dramatically suppressed as temperature decreases and at $U_{Mott}\thickapprox4.2$ one observes a sharp drop to zero, suggesting a Mott transition for a value of $U$ which is in good agreement with the position of the transition point shown on the phase diagram Fig. \ref{fig:phaseDiagram}, where
  the emerging finite single-particle gap was used to determine $U_{Mott}$.

\begin{figure}[!t]
\begin{center}
\includegraphics[width=9cm]{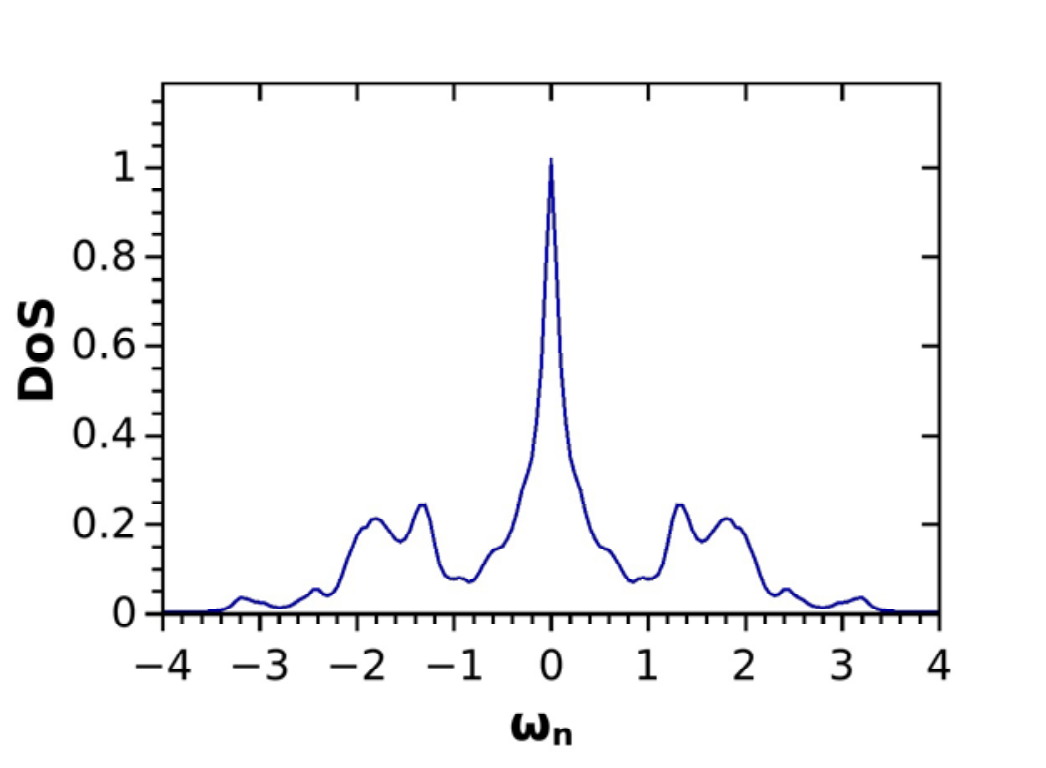}
\end{center}
\caption{(Color online) A sketch of the density of states of the conduction band of the effective Anderson impurity model obtained from DMFT on the honeycomb lattice.
300 bath sites are used to parameterize the effective Anderson impurity model. The spectrum of bath-level energies is displayed with a broadening factor $\delta = 0.08$.
}
\label{fig:conduction}
\end{figure}

We can gain insight into this apparent contradiction between the under-screened fixed point that exists in the Anderson impurity problem and our Fermi-liquid result  
by studying the basic difference between the effective cluster model of DCA and the pseudogap Anderson/Kondo model. 
Without loss of generality, we consider a single-site DMFT effective model \cite{georges1996dynamical} which can be parameterized 
as an impurity site coupled homogeneously to a conduction band. The DMFT self-consistent equation 
requires the local density of states on the impurity site to be just the same as that of the lattice model, 
thus on Fermi level it must hybridized strongly to the conduction band to obtain a semi-metallic behavior in the local density of states. 
As a result, the effective conduction band of the DMFT impurity model must have a large density of states at the Fermi level
(see Fig.\ref{fig:conduction})
which is in contrast to the pseudogap Anderson impurity model. Consequently, we conclude that the under-screened fixed point
scenario of pseudogap Anderson impurity model does not apply to the Hubbard model on the honeycomb lattice.

\begin{figure}[!t]
\includegraphics[width=9cm]{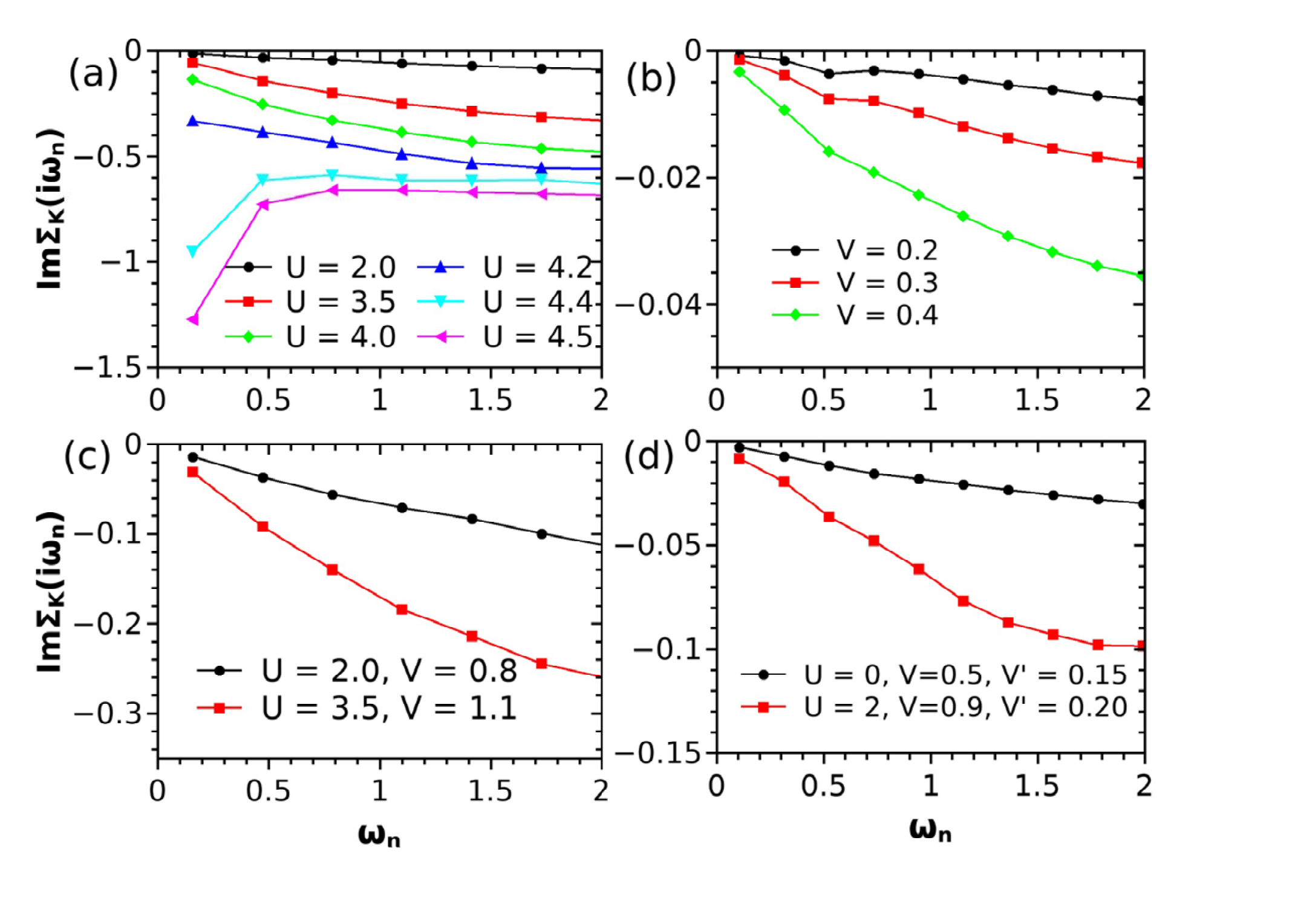}
\caption{(Color online) Imaginary part of the self-energies as a function of Matsubara frequency at the Dirac point $K$. (a) For the Hubbard model ($V =V'=0$) at $T=1/20$. 
(b) For various values of the nearest-neighbor repulsion $V$ at $T=1/30$ (c) For both on-site $U$ and nearest-neighbor repulsion $V$ at $T= 1/20$ 
(d) For on-site $U$, nearest-neighbor repulsion $V$ and next-nearest-neighbor repulsion $V'$ at  $T= 1/30$.
}
\label{fig:selfenergy}
\end{figure}

In order to find out whether the Fermi liquid in this many-body system can be broken down by correlation effects, 
we plot in Fig. \ref{fig:selfenergy} the imaginary part of the self-energy $\Sigma(i\omega_{n})$ at the Dirac point $K$ as
a function of Matsubara frequency.
Due to particle-hole symmetry, the real part of $\Sigma_{K}(i\omega_{n})$
  vanishes in our study while the imaginary part should scale linearly in the low-energy 
  region if the system preserves the Fermi-liquid behavior.  We investigate the effect of on-site $U$, nearest-neighbor $V$ and next-nearest-neighbor $V'$ interactions on the self-energy

  We begin with Fig. \ref{fig:selfenergy}a, where only $U$ differs from zero. In this case, the system
  is governed by spin fluctuations and hence competition between Kondo screening and RKKY correlations dominates the physical
  properties. This is reflected by the evolution of the self-energy, as we now show. As the 
  interaction increases, there is a gradual transition from the Fermi liquid ($U=2.0, 3.5, 4.0$) state to a crossover region ($U=4.2$), to a bad insulator ($U=4.4$) state and finally to a Mott insulator at $U=4.5$. The signature of the transition to the Mott insulating state is not as sharply defined as in our earlier approach. 
  
  To study the effects of the charge density
  fluctuation on the single-particle dynamics, we consider the effect of $V$ in Fig.~\ref{fig:selfenergy}b and Fig.~\ref{fig:selfenergy}c.
  No breakdown of Fermi-liquid behavior is observed for all of the parameters, as long as the SDW/CDW phase transitions are avoided. 
   From Fig.~\ref{fig:selfenergy}c we also see clearly that even in the presence of both $U$ and $V$, we recover a Fermi liquid behavior. We also checked for on-site Hubbard interaction, $U =2$ and $U=3.5$, that stronger inter-site interaction $V$  
   decreases the magnitude of the self-energy.  
  Note that
 when $U, V$ are weak in Figs.~\ref{fig:selfenergy}b,c one observes kinks on the self-energy caused by the charge-density fluctuations. These kinks eventually fade out
  when $U, V$ become strong. Fig. \ref{fig:selfenergy}d shows that when $V'$, namely next-nearest-neighbor repulsion, is turned on to destroy the CDW order,
the  Fermi liquid immediately recovers. 
  
  Finally, note that although the RPA calculations with full Coulomb potential suggest a marginal Fermi liquid (MFL) in graphene \cite{sarma2007many},
  a more sophisticated RG analysis shows that the MFL is ultimately avoided when the running of the interaction parameters
  is taken into account~\cite{gonzalez1999marginal, kotov2012electron}. This is in agreement  with our present result.

 \subsection{Fermi velocity and correlation effects}
\begin{figure}
\begin{center}
\includegraphics[width=9cm]{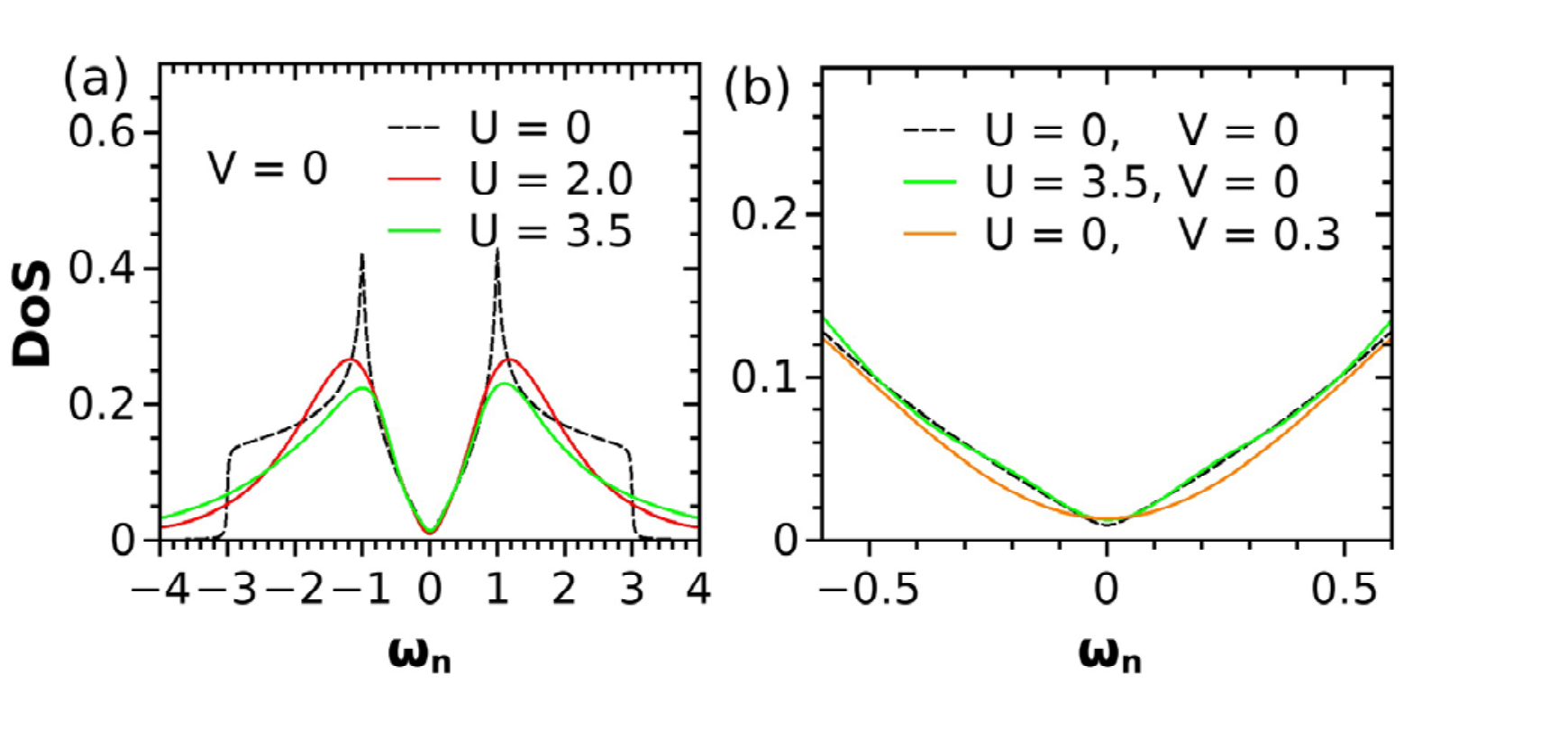}
\end{center}
\caption{(Color online) Local density of states (LDOS) at $T= 0.05$ (a) For the Hubbard model ($V =0$) with $U=2.0$ (red line) and $U=3.5$ (green line) obtained from analytical continuation using the maximum entropy method (MEM). The non-interacting case  is shown for reference with the the dashed black line.
(b) Comparison of the non-interacting LDOS (black line) near the Fermi level with results obtained with pure on-site repulsion $U$ (green line) and pure nearest-neighbor repulsion $V$ (red line). The Pad\'e approximation method is used to generate high precision LDOS data at low-energies.
}
\label{fig:dos}
\end{figure}
  In graphene the Fermi velocity $v_{F}$ is an important physical quantity which defines the low-energy effective theory of the Dirac fermions. 
  For the non-interacting case, the Fermi velocity $v_{F}$ is related to the local density of states (LDOS) via \cite{neto2009electronic},
\begin{equation}
 \rho(E)	=\frac{3\sqrt{3}a^{2}|E|}{\pi}\times\frac{1}{v_{F}^{2}}
\end{equation}
where $a$ is the lattice constant. The Fermi velocity $v_{F}$ is usually estimated as $10^{6} m/s$ for the tight-binding model.~\cite{wallace1947band} Recent studies ~\cite{elias2011dirac} suggest that the Fermi velocity is renormalized in suspended graphene. In this subsection, we investigate how interactions renormalize this velocity.

In order to obtain the LDOS in real frequency space, we employ the maximum entropy method (MEM) \cite{jarrell1996bayesian} and the Pad\'e approximation \cite{liebsch2013coulomb} 
to perform the numerical analytical continuation of the imaginary-frequency local Green's function $G(i\omega_{n})$.
In Fig. \ref{fig:dos}a, we first show the LDOS for DCA with only the on-site interaction. We observe that while the van Hove singularity is rounded and the high-energy LDOS decreases in magnitude, spreading beyond the non-interacting bandwidth, the LDOS in the vicinity of 
Fermi level stays constant as the interaction strength increases. In other words, when only the Hubbard interaction 
is considered, large cluster DCA calculations demonstrate that $v_{F}$ is insensitive to the interaction strength, sticking to the tight-binding model
value of $10^{6}m/s$. This is in agreement with the previous cellular DMFT (CDMFT) result \cite{wu2010interacting} and RG analysis \cite{herbut2009theory}.
We note that studies with DCA on smaller clusters~\cite{wu2010interacting} suggest that 
$v_{F}$ decreases as the Hubbard interaction increases. In contrast, CDMFT 
is able to find the correct invariable $v_{F}$ even in the case of small clusters \cite{wu2010interacting}. 
This observation is in agreement with previous benchmark calculations that suggest that on small clusters, CDMFT converges faster than DCA for local quantities such as the LDOS.~\cite{biroli2002cluster,Kyung2006} Limitations of DCA have been discussed in this context in Ref.~\onlinecite{liebsch2013coulomb}. 

In Fig. \ref{fig:dos}b, we compare the LDOS when only nearest-neighbor repulsion $V=0.3$ is taken into account
with the LDOS for $U = 3.5, V=0$. As we can see,
for $U=3.5$ the LDOS almost superposes with the non-interacting result at low-energies, whereas the $V=0.3$ curve is significantly below 
the  non-interacting result, suggesting an increased $v_{F}$. This dramatic difference between
the correlation effects of local and non-local repulsions on the single-particle dynamics leads us to conclude that the observed increase of $v_{F}$ in suspended graphene \cite{elias2011dirac} should be attributed to 
the long range Coulomb potential rather than to the local Hubbard repulsion.


\section{Summary and discussion}\label{Sec:Summary}
Using clusters as large as 96 sites with the Dynamical Cluster Approximation, we have found the phase diagram of correlated electrons on the honeycomb lattice at half-filling. SDW, CDW and Fermi-liquid semi-metallic phases are present when on-site interaction $U$ and nearest-neighbor repulsion $V$ are taken into account.  

As a function of temperature and on-site interaction $U$ (for $V=0$) the semi-metal turns into a state where the A and B sub-lattices order antiferromagnetically with respect to each other at low temperature when $U$ increases. At finite $T$, this antiferromagnetic transition should be interpreted as a crossover to the renormalized classical regime since in two dimensions antiferromagnetic long-range order can occur only at $T=0$, as required by the Mermin-Wagner theorem.  The zero-temperature Mott transition is masked by the antiferromagnetic phase so that there is no spin-liquid regime. 

At low temperature, the effect of a large enough near-neighbor repulsion $V$ is to induce a staggered CDW phase where A and B sublattices exhibit an excess of either electrons or holes, depending on the sublattice. The transition
line between the SDW and staggered CDW phases is roughly determined by $U\approx 3V$ when $U$ is significantly larger than the hopping amplitude $t$. The transition to the CDW phase is always first order but the semi-metallic to SDW transition is continuous. 

We have also investigated in detail the properties of the Fermi-liquid phase. As temperature
decreases, the high-temperature bad metal phase screens off the local moments that eventually evolve into a Fermi-liquid phase at low temperature, despite the semi-metallic nature of the density of states. We have shown that this occurs because the self-consistent bath is in fact metallic. This Fermi-liquid
state is stable even in the presence of non-local Coulomb interactions, at least up to next-nearest-neighbor repulsion $V'$, as long as the mutual competitions between repulsive potentials do not order the system. 

Moreover, we demonstrated that the short-ranged interaction $U$ does not lead to a measurable renormalization of the Fermi velocity $v_F$ close to Fermi level. However, $v_F$ is clearly renormalized in the presence of $V$. This could 
explain the variations of $v_{F}$ observed in different experiments on graphene. Since $v_F$ is not sensitive to the 
short-ranged repulsion, graphene placed in high dielectric-constant environment should have a Fermi velocity less effected by correlation effects 
than graphene in vacuum or on a substrate with small dielectric constant $\epsilon$, which is expected to have a strongly renormalized $v_F$. 

\acknowledgments

W. W. would like to thank Ansgar Liebsch for useful discussions about the parameterization algorithms of DCA. 
This work was supported by the Natural Sciences and Engineering Research Council of Canada (NSERC), and by the Tier I Canada Research Chair Program (A.-M.S.T.). Simulations were performed on computers provided by CFI, MELS, Calcul Qu\'ebec and Compute Canada.

\bibliography{graphene2013}

\begin{thebibliography}{40}%
\makeatletter
\providecommand \@ifxundefined [1]{%
 \@ifx{#1\undefined}
}%
\providecommand \@ifnum [1]{%
 \ifnum #1\expandafter \@firstoftwo
 \else \expandafter \@secondoftwo
 \fi
}%
\providecommand \@ifx [1]{%
 \ifx #1\expandafter \@firstoftwo
 \else \expandafter \@secondoftwo
 \fi
}%
\providecommand \natexlab [1]{#1}%
\providecommand \enquote  [1]{``#1''}%
\providecommand \bibnamefont  [1]{#1}%
\providecommand \bibfnamefont [1]{#1}%
\providecommand \citenamefont [1]{#1}%
\providecommand \href@noop [0]{\@secondoftwo}%
\providecommand \href [0]{\begingroup \@sanitize@url \@href}%
\providecommand \@href[1]{\@@startlink{#1}\@@href}%
\providecommand \@@href[1]{\endgroup#1\@@endlink}%
\providecommand \@sanitize@url [0]{\catcode `\\12\catcode `\$12\catcode
  `\&12\catcode `\#12\catcode `\^12\catcode `\_12\catcode `\%12\relax}%
\providecommand \@@startlink[1]{}%
\providecommand \@@endlink[0]{}%
\providecommand \url  [0]{\begingroup\@sanitize@url \@url }%
\providecommand \@url [1]{\endgroup\@href {#1}{\urlprefix }}%
\providecommand \urlprefix  [0]{URL }%
\providecommand \Eprint [0]{\href }%
\providecommand \doibase [0]{http://dx.doi.org/}%
\providecommand \selectlanguage [0]{\@gobble}%
\providecommand \bibinfo  [0]{\@secondoftwo}%
\providecommand \bibfield  [0]{\@secondoftwo}%
\providecommand \translation [1]{[#1]}%
\providecommand \BibitemOpen [0]{}%
\providecommand \bibitemStop [0]{}%
\providecommand \bibitemNoStop [0]{.\EOS\space}%
\providecommand \EOS [0]{\spacefactor3000\relax}%
\providecommand \BibitemShut  [1]{\csname bibitem#1\endcsname}%
\let\auto@bib@innerbib\@empty
\bibitem [{\citenamefont {Castro~Neto}\ \emph {et~al.}(2009)\citenamefont
  {Castro~Neto}, \citenamefont {Guinea}, \citenamefont {Peres}, \citenamefont
  {Novoselov},\ and\ \citenamefont {Geim}}]{neto2009electronic}%
  \BibitemOpen
  \bibfield  {author} {\bibinfo {author} {\bibfnamefont {A.~H.}\ \bibnamefont
  {Castro~Neto}}, \bibinfo {author} {\bibfnamefont {F.}~\bibnamefont {Guinea}},
  \bibinfo {author} {\bibfnamefont {N.~M.~R.}\ \bibnamefont {Peres}}, \bibinfo
  {author} {\bibfnamefont {K.~S.}\ \bibnamefont {Novoselov}}, \ and\ \bibinfo
  {author} {\bibfnamefont {A.~K.}\ \bibnamefont {Geim}},\ }\href {\doibase
  10.1103/RevModPhys.81.109} {\bibfield  {journal} {\bibinfo  {journal} {Rev.
  Mod. Phys.}\ }\textbf {\bibinfo {volume} {81}},\ \bibinfo {pages} {109}
  (\bibinfo {year} {2009})}\BibitemShut {NoStop}%
\bibitem [{\citenamefont {Elias}\ \emph {et~al.}(2011)\citenamefont {Elias},
  \citenamefont {Gorbachev}, \citenamefont {Mayorov}, \citenamefont {Morozov},
  \citenamefont {Zhukov}, \citenamefont {Blake}, \citenamefont {Grigorieva},
  \citenamefont {Novoselov}, \citenamefont {Guinea},\ and\ \citenamefont
  {Geim}}]{elias2011dirac}%
  \BibitemOpen
  \bibfield  {author} {\bibinfo {author} {\bibfnamefont {D.}~\bibnamefont
  {Elias}}, \bibinfo {author} {\bibfnamefont {R.}~\bibnamefont {Gorbachev}},
  \bibinfo {author} {\bibfnamefont {A.}~\bibnamefont {Mayorov}}, \bibinfo
  {author} {\bibfnamefont {S.}~\bibnamefont {Morozov}}, \bibinfo {author}
  {\bibfnamefont {A.}~\bibnamefont {Zhukov}}, \bibinfo {author} {\bibfnamefont
  {P.}~\bibnamefont {Blake}}, \bibinfo {author} {\bibfnamefont {L.~P.~I.}\
  \bibnamefont {Grigorieva}}, \bibinfo {author} {\bibfnamefont
  {K.}~\bibnamefont {Novoselov}}, \bibinfo {author} {\bibfnamefont
  {F.}~\bibnamefont {Guinea}}, \ and\ \bibinfo {author} {\bibfnamefont
  {A.}~\bibnamefont {Geim}},\ }\href
  {http://www.nature.com/nphys/journal/v7/n9/full/nphys2049.html} {\bibfield
  {journal} {\bibinfo  {journal} {Nature Physics}\ }\textbf {\bibinfo {volume}
  {7}},\ \bibinfo {pages} {701} (\bibinfo {year} {2011})}\BibitemShut {NoStop}%
\bibitem [{\citenamefont {Das~Sarma}\ \emph {et~al.}(2007)\citenamefont
  {Das~Sarma}, \citenamefont {Hwang},\ and\ \citenamefont
  {Tse}}]{sarma2007many}%
  \BibitemOpen
  \bibfield  {author} {\bibinfo {author} {\bibfnamefont {S.}~\bibnamefont
  {Das~Sarma}}, \bibinfo {author} {\bibfnamefont {E.~H.}\ \bibnamefont
  {Hwang}}, \ and\ \bibinfo {author} {\bibfnamefont {W.-K.}\ \bibnamefont
  {Tse}},\ }\href {\doibase 10.1103/PhysRevB.75.121406} {\bibfield  {journal}
  {\bibinfo  {journal} {Phys. Rev. B}\ }\textbf {\bibinfo {volume} {75}},\
  \bibinfo {pages} {121406} (\bibinfo {year} {2007})}\BibitemShut {NoStop}%
\bibitem [{\citenamefont {Wehling}\ \emph {et~al.}(2011)\citenamefont
  {Wehling}, \citenamefont {\ifmmode \mbox{\c{S}}\else \c{S}\fi{}a\ifmmode
  \mbox{\c{s}}\else \c{s}\fi{}\ifmmode \imath \else \i
  \fi{}o\ifmmode~\breve{g}\else \u{g}\fi{}lu}, \citenamefont {Friedrich},
  \citenamefont {Lichtenstein}, \citenamefont {Katsnelson},\ and\ \citenamefont
  {Bl\"ugel}}]{wehling2011strength}%
  \BibitemOpen
  \bibfield  {author} {\bibinfo {author} {\bibfnamefont {T.~O.}\ \bibnamefont
  {Wehling}}, \bibinfo {author} {\bibfnamefont {E.}~\bibnamefont {\ifmmode
  \mbox{\c{S}}\else \c{S}\fi{}a\ifmmode \mbox{\c{s}}\else \c{s}\fi{}\ifmmode
  \imath \else \i \fi{}o\ifmmode~\breve{g}\else \u{g}\fi{}lu}}, \bibinfo
  {author} {\bibfnamefont {C.}~\bibnamefont {Friedrich}}, \bibinfo {author}
  {\bibfnamefont {A.~I.}\ \bibnamefont {Lichtenstein}}, \bibinfo {author}
  {\bibfnamefont {M.~I.}\ \bibnamefont {Katsnelson}}, \ and\ \bibinfo {author}
  {\bibfnamefont {S.}~\bibnamefont {Bl\"ugel}},\ }\href {\doibase
  10.1103/PhysRevLett.106.236805} {\bibfield  {journal} {\bibinfo  {journal}
  {Phys. Rev. Lett.}\ }\textbf {\bibinfo {volume} {106}},\ \bibinfo {pages}
  {236805} (\bibinfo {year} {2011})}\BibitemShut {NoStop}%
\bibitem [{\citenamefont {Chae}\ \emph {et~al.}(2012)\citenamefont {Chae},
  \citenamefont {Jung}, \citenamefont {Young}, \citenamefont {Dean},
  \citenamefont {Wang}, \citenamefont {Gao}, \citenamefont {Watanabe},
  \citenamefont {Taniguchi}, \citenamefont {Hone}, \citenamefont {Shepard},
  \citenamefont {Kim}, \citenamefont {Zhitenev},\ and\ \citenamefont
  {Stroscio}}]{chae2012renormalization}%
  \BibitemOpen
  \bibfield  {author} {\bibinfo {author} {\bibfnamefont {J.}~\bibnamefont
  {Chae}}, \bibinfo {author} {\bibfnamefont {S.}~\bibnamefont {Jung}}, \bibinfo
  {author} {\bibfnamefont {A.~F.}\ \bibnamefont {Young}}, \bibinfo {author}
  {\bibfnamefont {C.~R.}\ \bibnamefont {Dean}}, \bibinfo {author}
  {\bibfnamefont {L.}~\bibnamefont {Wang}}, \bibinfo {author} {\bibfnamefont
  {Y.}~\bibnamefont {Gao}}, \bibinfo {author} {\bibfnamefont {K.}~\bibnamefont
  {Watanabe}}, \bibinfo {author} {\bibfnamefont {T.}~\bibnamefont {Taniguchi}},
  \bibinfo {author} {\bibfnamefont {J.}~\bibnamefont {Hone}}, \bibinfo {author}
  {\bibfnamefont {K.~L.}\ \bibnamefont {Shepard}}, \bibinfo {author}
  {\bibfnamefont {P.}~\bibnamefont {Kim}}, \bibinfo {author} {\bibfnamefont
  {N.~B.}\ \bibnamefont {Zhitenev}}, \ and\ \bibinfo {author} {\bibfnamefont
  {J.~A.}\ \bibnamefont {Stroscio}},\ }\href {\doibase
  10.1103/PhysRevLett.109.116802} {\bibfield  {journal} {\bibinfo  {journal}
  {Phys. Rev. Lett.}\ }\textbf {\bibinfo {volume} {109}},\ \bibinfo {pages}
  {116802} (\bibinfo {year} {2012})}\BibitemShut {NoStop}%
\bibitem [{\citenamefont {Martin}\ \emph {et~al.}(2007)\citenamefont {Martin},
  \citenamefont {Akerman}, \citenamefont {Ulbricht}, \citenamefont {Lohmann},
  \citenamefont {Smet}, \citenamefont {Von~Klitzing},\ and\ \citenamefont
  {Yacoby}}]{martin2007observation}%
  \BibitemOpen
  \bibfield  {author} {\bibinfo {author} {\bibfnamefont {J.}~\bibnamefont
  {Martin}}, \bibinfo {author} {\bibfnamefont {N.}~\bibnamefont {Akerman}},
  \bibinfo {author} {\bibfnamefont {G.}~\bibnamefont {Ulbricht}}, \bibinfo
  {author} {\bibfnamefont {T.}~\bibnamefont {Lohmann}}, \bibinfo {author}
  {\bibfnamefont {J.}~\bibnamefont {Smet}}, \bibinfo {author} {\bibfnamefont
  {K.}~\bibnamefont {Von~Klitzing}}, \ and\ \bibinfo {author} {\bibfnamefont
  {A.}~\bibnamefont {Yacoby}},\ }\href
  {http://www.nature.com/nphys/journal/v4/n2/full/nphys781.html} {\bibfield
  {journal} {\bibinfo  {journal} {Nature Physics}\ }\textbf {\bibinfo {volume}
  {4}},\ \bibinfo {pages} {144} (\bibinfo {year} {2007})}\BibitemShut {NoStop}%
\bibitem [{\citenamefont {Yankowitz}\ \emph {et~al.}(2012)\citenamefont
  {Yankowitz}, \citenamefont {Xue}, \citenamefont {Cormode}, \citenamefont
  {Sanchez-Yamagishi}, \citenamefont {Watanabe}, \citenamefont {Taniguchi},
  \citenamefont {Jarillo-Herrero}, \citenamefont {Jacquod},\ and\ \citenamefont
  {LeRoy}}]{yankowitz2012emergence}%
  \BibitemOpen
  \bibfield  {author} {\bibinfo {author} {\bibfnamefont {M.}~\bibnamefont
  {Yankowitz}}, \bibinfo {author} {\bibfnamefont {J.}~\bibnamefont {Xue}},
  \bibinfo {author} {\bibfnamefont {D.}~\bibnamefont {Cormode}}, \bibinfo
  {author} {\bibfnamefont {J.~D.}\ \bibnamefont {Sanchez-Yamagishi}}, \bibinfo
  {author} {\bibfnamefont {K.}~\bibnamefont {Watanabe}}, \bibinfo {author}
  {\bibfnamefont {T.}~\bibnamefont {Taniguchi}}, \bibinfo {author}
  {\bibfnamefont {P.}~\bibnamefont {Jarillo-Herrero}}, \bibinfo {author}
  {\bibfnamefont {P.}~\bibnamefont {Jacquod}}, \ and\ \bibinfo {author}
  {\bibfnamefont {B.~J.}\ \bibnamefont {LeRoy}},\ }\href
  {http://www.nature.com/nphys/journal/v8/n5/full/nphys2272.html} {\bibfield
  {journal} {\bibinfo  {journal} {Nature Physics}\ }\textbf {\bibinfo {volume}
  {8}},\ \bibinfo {pages} {382} (\bibinfo {year} {2012})}\BibitemShut {NoStop}%
\bibitem [{\citenamefont {Reed}\ \emph {et~al.}(2010)\citenamefont {Reed},
  \citenamefont {Uchoa}, \citenamefont {Joe}, \citenamefont {Gan},
  \citenamefont {Casa}, \citenamefont {Fradkin},\ and\ \citenamefont
  {Abbamonte}}]{reed2010effective}%
  \BibitemOpen
  \bibfield  {author} {\bibinfo {author} {\bibfnamefont {J.~P.}\ \bibnamefont
  {Reed}}, \bibinfo {author} {\bibfnamefont {B.}~\bibnamefont {Uchoa}},
  \bibinfo {author} {\bibfnamefont {Y.~I.}\ \bibnamefont {Joe}}, \bibinfo
  {author} {\bibfnamefont {Y.}~\bibnamefont {Gan}}, \bibinfo {author}
  {\bibfnamefont {D.}~\bibnamefont {Casa}}, \bibinfo {author} {\bibfnamefont
  {E.}~\bibnamefont {Fradkin}}, \ and\ \bibinfo {author} {\bibfnamefont
  {P.}~\bibnamefont {Abbamonte}},\ }\href
  {http://www.sciencemag.org/content/330/6005/805.full} {\bibfield  {journal}
  {\bibinfo  {journal} {Science}\ }\textbf {\bibinfo {volume} {330}},\ \bibinfo
  {pages} {805} (\bibinfo {year} {2010})}\BibitemShut {NoStop}%
\bibitem [{\citenamefont {Sprinkle}\ \emph {et~al.}(2009)\citenamefont
  {Sprinkle}, \citenamefont {Siegel}, \citenamefont {Hu}, \citenamefont
  {Hicks}, \citenamefont {Tejeda}, \citenamefont {Taleb-Ibrahimi},
  \citenamefont {Le~F\`evre}, \citenamefont {Bertran}, \citenamefont {Vizzini},
  \citenamefont {Enriquez}, \citenamefont {Chiang}, \citenamefont
  {Soukiassian}, \citenamefont {Berger}, \citenamefont {de~Heer}, \citenamefont
  {Lanzara},\ and\ \citenamefont {Conrad}}]{sprinkle2009first}%
  \BibitemOpen
  \bibfield  {author} {\bibinfo {author} {\bibfnamefont {M.}~\bibnamefont
  {Sprinkle}}, \bibinfo {author} {\bibfnamefont {D.}~\bibnamefont {Siegel}},
  \bibinfo {author} {\bibfnamefont {Y.}~\bibnamefont {Hu}}, \bibinfo {author}
  {\bibfnamefont {J.}~\bibnamefont {Hicks}}, \bibinfo {author} {\bibfnamefont
  {A.}~\bibnamefont {Tejeda}}, \bibinfo {author} {\bibfnamefont
  {A.}~\bibnamefont {Taleb-Ibrahimi}}, \bibinfo {author} {\bibfnamefont
  {P.}~\bibnamefont {Le~F\`evre}}, \bibinfo {author} {\bibfnamefont
  {F.}~\bibnamefont {Bertran}}, \bibinfo {author} {\bibfnamefont
  {S.}~\bibnamefont {Vizzini}}, \bibinfo {author} {\bibfnamefont
  {H.}~\bibnamefont {Enriquez}}, \bibinfo {author} {\bibfnamefont
  {S.}~\bibnamefont {Chiang}}, \bibinfo {author} {\bibfnamefont
  {P.}~\bibnamefont {Soukiassian}}, \bibinfo {author} {\bibfnamefont
  {C.}~\bibnamefont {Berger}}, \bibinfo {author} {\bibfnamefont {W.~A.}\
  \bibnamefont {de~Heer}}, \bibinfo {author} {\bibfnamefont {A.}~\bibnamefont
  {Lanzara}}, \ and\ \bibinfo {author} {\bibfnamefont {E.~H.}\ \bibnamefont
  {Conrad}},\ }\href {\doibase 10.1103/PhysRevLett.103.226803} {\bibfield
  {journal} {\bibinfo  {journal} {Phys. Rev. Lett.}\ }\textbf {\bibinfo
  {volume} {103}},\ \bibinfo {pages} {226803} (\bibinfo {year}
  {2009})}\BibitemShut {NoStop}%
\bibitem [{\citenamefont {Wu}\ \emph {et~al.}(2010)\citenamefont {Wu},
  \citenamefont {Chen}, \citenamefont {Tao}, \citenamefont {Tong},\ and\
  \citenamefont {Liu}}]{wu2010interacting}%
  \BibitemOpen
  \bibfield  {author} {\bibinfo {author} {\bibfnamefont {W.}~\bibnamefont
  {Wu}}, \bibinfo {author} {\bibfnamefont {Y.-H.}\ \bibnamefont {Chen}},
  \bibinfo {author} {\bibfnamefont {H.-S.}\ \bibnamefont {Tao}}, \bibinfo
  {author} {\bibfnamefont {N.-H.}\ \bibnamefont {Tong}}, \ and\ \bibinfo
  {author} {\bibfnamefont {W.-M.}\ \bibnamefont {Liu}},\ }\href {\doibase
  10.1103/PhysRevB.82.245102} {\bibfield  {journal} {\bibinfo  {journal} {Phys.
  Rev. B}\ }\textbf {\bibinfo {volume} {82}},\ \bibinfo {pages} {245102}
  (\bibinfo {year} {2010})}\BibitemShut {NoStop}%
\bibitem [{\citenamefont {Liebsch}\ and\ \citenamefont
  {Wu}(2013)}]{liebsch2013coulomb}%
  \BibitemOpen
  \bibfield  {author} {\bibinfo {author} {\bibfnamefont {A.}~\bibnamefont
  {Liebsch}}\ and\ \bibinfo {author} {\bibfnamefont {W.}~\bibnamefont {Wu}},\
  }\href {\doibase 10.1103/PhysRevB.87.205127} {\bibfield  {journal} {\bibinfo
  {journal} {Phys. Rev. B}\ }\textbf {\bibinfo {volume} {87}},\ \bibinfo
  {pages} {205127} (\bibinfo {year} {2013})}\BibitemShut {NoStop}%
\bibitem [{\citenamefont {Meng}\ \emph {et~al.}(2010)\citenamefont {Meng},
  \citenamefont {Lang}, \citenamefont {Wessel}, \citenamefont {Assaad},\ and\
  \citenamefont {Muramatsu}}]{meng2010quantum}%
  \BibitemOpen
  \bibfield  {author} {\bibinfo {author} {\bibfnamefont {Z.}~\bibnamefont
  {Meng}}, \bibinfo {author} {\bibfnamefont {T.}~\bibnamefont {Lang}}, \bibinfo
  {author} {\bibfnamefont {S.}~\bibnamefont {Wessel}}, \bibinfo {author}
  {\bibfnamefont {F.}~\bibnamefont {Assaad}}, \ and\ \bibinfo {author}
  {\bibfnamefont {A.}~\bibnamefont {Muramatsu}},\ }\href
  {http://www.nature.com/nature/journal/v464/n7290/full/nature08942.html}
  {\bibfield  {journal} {\bibinfo  {journal} {Nature}\ }\textbf {\bibinfo
  {volume} {464}},\ \bibinfo {pages} {847} (\bibinfo {year}
  {2010})}\BibitemShut {NoStop}%
\bibitem [{\citenamefont {Assaad}\ and\ \citenamefont
  {Herbut}(2013)}]{assaad2013}%
  \BibitemOpen
  \bibfield  {author} {\bibinfo {author} {\bibfnamefont {F.~F.}\ \bibnamefont
  {Assaad}}\ and\ \bibinfo {author} {\bibfnamefont {I.~F.}\ \bibnamefont
  {Herbut}},\ }\href {\doibase 10.1103/PhysRevX.3.031010} {\bibfield  {journal}
  {\bibinfo  {journal} {Phys. Rev. X}\ }\textbf {\bibinfo {volume} {3}},\
  \bibinfo {pages} {031010} (\bibinfo {year} {2013})}\BibitemShut {NoStop}%
\bibitem [{\citenamefont {Sorella}\ \emph {et~al.}(2012)\citenamefont
  {Sorella}, \citenamefont {Otsuka},\ and\ \citenamefont
  {Yunoki}}]{sorella2012absence}%
  \BibitemOpen
  \bibfield  {author} {\bibinfo {author} {\bibfnamefont {S.}~\bibnamefont
  {Sorella}}, \bibinfo {author} {\bibfnamefont {Y.}~\bibnamefont {Otsuka}}, \
  and\ \bibinfo {author} {\bibfnamefont {S.}~\bibnamefont {Yunoki}},\ }\href
  {http://www.nature.com/srep/2012/121218/srep00992/full/srep00992.html}
  {\bibfield  {journal} {\bibinfo  {journal} {Scientific reports}\ }\textbf
  {\bibinfo {volume} {2}} (\bibinfo {year} {2012})}\BibitemShut {NoStop}%
\bibitem [{\citenamefont {Blankenbecler}\ \emph {et~al.}(1981)\citenamefont
  {Blankenbecler}, \citenamefont {Scalapino},\ and\ \citenamefont
  {Sugar}}]{blankenbecler1981monte}%
  \BibitemOpen
  \bibfield  {author} {\bibinfo {author} {\bibfnamefont {R.}~\bibnamefont
  {Blankenbecler}}, \bibinfo {author} {\bibfnamefont {D.~J.}\ \bibnamefont
  {Scalapino}}, \ and\ \bibinfo {author} {\bibfnamefont {R.~L.}\ \bibnamefont
  {Sugar}},\ }\href {\doibase 10.1103/PhysRevD.24.2278} {\bibfield  {journal}
  {\bibinfo  {journal} {Phys. Rev. D}\ }\textbf {\bibinfo {volume} {24}},\
  \bibinfo {pages} {2278} (\bibinfo {year} {1981})}\BibitemShut {NoStop}%
\bibitem [{\citenamefont {Loh~Jr}\ and\ \citenamefont
  {Gubernatis}(1992)}]{loh1992stable}%
  \BibitemOpen
  \bibfield  {author} {\bibinfo {author} {\bibfnamefont {E.}~\bibnamefont
  {Loh~Jr}}\ and\ \bibinfo {author} {\bibfnamefont {J.}~\bibnamefont
  {Gubernatis}},\ }\href@noop {} {\bibfield  {journal} {\bibinfo  {journal}
  {Modern Problems of Condensed Matter Physics}\ }\textbf {\bibinfo {volume}
  {32}},\ \bibinfo {pages} {177} (\bibinfo {year} {1992})}\BibitemShut
  {NoStop}%
\bibitem [{\citenamefont {Wallace}(1947)}]{wallace1947band}%
  \BibitemOpen
  \bibfield  {author} {\bibinfo {author} {\bibfnamefont {P.~R.}\ \bibnamefont
  {Wallace}},\ }\href {\doibase 10.1103/PhysRev.71.622} {\bibfield  {journal}
  {\bibinfo  {journal} {Phys. Rev.}\ }\textbf {\bibinfo {volume} {71}},\
  \bibinfo {pages} {622} (\bibinfo {year} {1947})}\BibitemShut {NoStop}%
\bibitem [{\citenamefont {Hettler}\ \emph {et~al.}(1998)\citenamefont
  {Hettler}, \citenamefont {Tahvildar-Zadeh}, \citenamefont {Jarrell},
  \citenamefont {Pruschke},\ and\ \citenamefont
  {Krishnamurthy}}]{Hettler:1998}%
  \BibitemOpen
  \bibfield  {author} {\bibinfo {author} {\bibfnamefont {M.~H.}\ \bibnamefont
  {Hettler}}, \bibinfo {author} {\bibfnamefont {A.~N.}\ \bibnamefont
  {Tahvildar-Zadeh}}, \bibinfo {author} {\bibfnamefont {M.}~\bibnamefont
  {Jarrell}}, \bibinfo {author} {\bibfnamefont {T.}~\bibnamefont {Pruschke}}, \
  and\ \bibinfo {author} {\bibfnamefont {H.~R.}\ \bibnamefont
  {Krishnamurthy}},\ }\href {\doibase 10.1103/PhysRevB.58.R7475} {\bibfield
  {journal} {\bibinfo  {journal} {Phys. Rev. B}\ }\textbf {\bibinfo {volume}
  {58}},\ \bibinfo {pages} {R7475} (\bibinfo {year} {1998})}\BibitemShut
  {NoStop}%
\bibitem [{\citenamefont {Khatami}\ \emph {et~al.}(2010)\citenamefont
  {Khatami}, \citenamefont {Lee}, \citenamefont {Bai}, \citenamefont
  {Scalettar},\ and\ \citenamefont {Jarrell}}]{KhatamiSolver:2010}%
  \BibitemOpen
  \bibfield  {author} {\bibinfo {author} {\bibfnamefont {E.}~\bibnamefont
  {Khatami}}, \bibinfo {author} {\bibfnamefont {C.~R.}\ \bibnamefont {Lee}},
  \bibinfo {author} {\bibfnamefont {Z.~J.}\ \bibnamefont {Bai}}, \bibinfo
  {author} {\bibfnamefont {R.~T.}\ \bibnamefont {Scalettar}}, \ and\ \bibinfo
  {author} {\bibfnamefont {M.}~\bibnamefont {Jarrell}},\ }\href {\doibase
  10.1103/PhysRevE.81.056703} {\bibfield  {journal} {\bibinfo  {journal} {Phys.
  Rev. E}\ }\textbf {\bibinfo {volume} {81}},\ \bibinfo {pages} {056703}
  (\bibinfo {year} {2010})}\BibitemShut {NoStop}%
\bibitem [{\citenamefont {Rubtsov}\ \emph {et~al.}(2005)\citenamefont
  {Rubtsov}, \citenamefont {Savkin},\ and\ \citenamefont
  {Lichtenstein}}]{rubtsov2005continuous}%
  \BibitemOpen
  \bibfield  {author} {\bibinfo {author} {\bibfnamefont {A.~N.}\ \bibnamefont
  {Rubtsov}}, \bibinfo {author} {\bibfnamefont {V.~V.}\ \bibnamefont {Savkin}},
  \ and\ \bibinfo {author} {\bibfnamefont {A.~I.}\ \bibnamefont
  {Lichtenstein}},\ }\href {\doibase 10.1103/PhysRevB.72.035122} {\bibfield
  {journal} {\bibinfo  {journal} {Phys. Rev. B}\ }\textbf {\bibinfo {volume}
  {72}},\ \bibinfo {pages} {035122} (\bibinfo {year} {2005})}\BibitemShut
  {NoStop}%
\bibitem [{\citenamefont {Park}\ \emph {et~al.}(2008)\citenamefont {Park},
  \citenamefont {Haule},\ and\ \citenamefont {Kotliar}}]{park2008cluster}%
  \BibitemOpen
  \bibfield  {author} {\bibinfo {author} {\bibfnamefont {H.}~\bibnamefont
  {Park}}, \bibinfo {author} {\bibfnamefont {K.}~\bibnamefont {Haule}}, \ and\
  \bibinfo {author} {\bibfnamefont {G.}~\bibnamefont {Kotliar}},\ }\href
  {\doibase 10.1103/PhysRevLett.101.186403} {\bibfield  {journal} {\bibinfo
  {journal} {Phys. Rev. Lett.}\ }\textbf {\bibinfo {volume} {101}},\ \bibinfo
  {pages} {186403} (\bibinfo {year} {2008})}\BibitemShut {NoStop}%
\bibitem [{\citenamefont {Arya}\ \emph {et~al.}(2014)\citenamefont {Arya},
  \citenamefont {Sriluckshmy}, \citenamefont {Hassan}, ,\ and\ \citenamefont
  {Tremblay}}]{Arya:2014}%
  \BibitemOpen
  \bibfield  {author} {\bibinfo {author} {\bibfnamefont {S.}~\bibnamefont
  {Arya}}, \bibinfo {author} {\bibfnamefont {P.}~\bibnamefont {Sriluckshmy}},
  \bibinfo {author} {\bibfnamefont {S.}~\bibnamefont {Hassan}}, , \ and\
  \bibinfo {author} {\bibfnamefont {A.-M.~S.}\ \bibnamefont {Tremblay}},\
  }\href@noop {} {\bibfield  {journal} {\bibinfo  {journal} {Unpublished}\ }
  (\bibinfo {year} {2014})}\BibitemShut {NoStop}%
\bibitem [{\citenamefont {Hassan}\ and\ \citenamefont
  {S\'en\'echal}(2013)}]{hassan2013}%
  \BibitemOpen
  \bibfield  {author} {\bibinfo {author} {\bibfnamefont {S.~R.}\ \bibnamefont
  {Hassan}}\ and\ \bibinfo {author} {\bibfnamefont {D.}~\bibnamefont
  {S\'en\'echal}},\ }\href {\doibase 10.1103/PhysRevLett.110.096402} {\bibfield
   {journal} {\bibinfo  {journal} {Phys. Rev. Lett.}\ }\textbf {\bibinfo
  {volume} {110}},\ \bibinfo {pages} {096402} (\bibinfo {year}
  {2013})}\BibitemShut {NoStop}%
\bibitem [{\citenamefont {Wu}\ \emph {et~al.}(2012)\citenamefont {Wu},
  \citenamefont {Rachel}, \citenamefont {Liu},\ and\ \citenamefont
  {Le~Hur}}]{WuLeHur:2012}%
  \BibitemOpen
  \bibfield  {author} {\bibinfo {author} {\bibfnamefont {W.}~\bibnamefont
  {Wu}}, \bibinfo {author} {\bibfnamefont {S.}~\bibnamefont {Rachel}}, \bibinfo
  {author} {\bibfnamefont {W.-M.}\ \bibnamefont {Liu}}, \ and\ \bibinfo
  {author} {\bibfnamefont {K.}~\bibnamefont {Le~Hur}},\ }\href {\doibase
  10.1103/PhysRevB.85.205102} {\bibfield  {journal} {\bibinfo  {journal} {Phys.
  Rev. B}\ }\textbf {\bibinfo {volume} {85}},\ \bibinfo {pages} {205102}
  (\bibinfo {year} {2012})}\BibitemShut {NoStop}%
\bibitem [{\citenamefont {Zhang}\ and\ \citenamefont
  {Callaway}(1989)}]{zhang1989extended}%
  \BibitemOpen
  \bibfield  {author} {\bibinfo {author} {\bibfnamefont {Y.}~\bibnamefont
  {Zhang}}\ and\ \bibinfo {author} {\bibfnamefont {J.}~\bibnamefont
  {Callaway}},\ }\href {\doibase 10.1103/PhysRevB.39.9397} {\bibfield
  {journal} {\bibinfo  {journal} {Phys. Rev. B}\ }\textbf {\bibinfo {volume}
  {39}},\ \bibinfo {pages} {9397} (\bibinfo {year} {1989})}\BibitemShut
  {NoStop}%
\bibitem [{\citenamefont {Hirsch}(1984)}]{hirsch1984charge}%
  \BibitemOpen
  \bibfield  {author} {\bibinfo {author} {\bibfnamefont {J.~E.}\ \bibnamefont
  {Hirsch}},\ }\href {\doibase 10.1103/PhysRevLett.53.2327} {\bibfield
  {journal} {\bibinfo  {journal} {Phys. Rev. Lett.}\ }\textbf {\bibinfo
  {volume} {53}},\ \bibinfo {pages} {2327} (\bibinfo {year}
  {1984})}\BibitemShut {NoStop}%
\bibitem [{\citenamefont {Aichhorn}\ \emph {et~al.}(2004)\citenamefont
  {Aichhorn}, \citenamefont {Evertz}, \citenamefont {von~der Linden},\ and\
  \citenamefont {Potthoff}}]{aichhorn2004charge}%
  \BibitemOpen
  \bibfield  {author} {\bibinfo {author} {\bibfnamefont {M.}~\bibnamefont
  {Aichhorn}}, \bibinfo {author} {\bibfnamefont {H.~G.}\ \bibnamefont
  {Evertz}}, \bibinfo {author} {\bibfnamefont {W.}~\bibnamefont {von~der
  Linden}}, \ and\ \bibinfo {author} {\bibfnamefont {M.}~\bibnamefont
  {Potthoff}},\ }\href {\doibase 10.1103/PhysRevB.70.235107} {\bibfield
  {journal} {\bibinfo  {journal} {Phys. Rev. B}\ }\textbf {\bibinfo {volume}
  {70}},\ \bibinfo {pages} {235107} (\bibinfo {year} {2004})}\BibitemShut
  {NoStop}%
\bibitem [{\citenamefont {Fourcade}\ and\ \citenamefont
  {Spronken}(1984)}]{Fourcade:1984}%
  \BibitemOpen
  \bibfield  {author} {\bibinfo {author} {\bibfnamefont {B.}~\bibnamefont
  {Fourcade}}\ and\ \bibinfo {author} {\bibfnamefont {G.}~\bibnamefont
  {Spronken}},\ }\href {\doibase 10.1103/PhysRevB.29.5089} {\bibfield
  {journal} {\bibinfo  {journal} {Phys. Rev. B}\ }\textbf {\bibinfo {volume}
  {29}},\ \bibinfo {pages} {5089} (\bibinfo {year} {1984})}\BibitemShut
  {NoStop}%
\bibitem [{\citenamefont {Raghu}\ \emph {et~al.}(2008)\citenamefont {Raghu},
  \citenamefont {Qi}, \citenamefont {Honerkamp},\ and\ \citenamefont
  {Zhang}}]{raghu_topological_2008}%
  \BibitemOpen
  \bibfield  {author} {\bibinfo {author} {\bibfnamefont {S.}~\bibnamefont
  {Raghu}}, \bibinfo {author} {\bibfnamefont {X.-L.}\ \bibnamefont {Qi}},
  \bibinfo {author} {\bibfnamefont {C.}~\bibnamefont {Honerkamp}}, \ and\
  \bibinfo {author} {\bibfnamefont {S.-C.}\ \bibnamefont {Zhang}},\ }\href
  {\doibase 10.1103/PhysRevLett.100.156401} {\bibfield  {journal} {\bibinfo
  {journal} {Physical Review Letters}\ }\textbf {\bibinfo {volume} {100}},\
  \bibinfo {pages} {156401} (\bibinfo {year} {2008})}\BibitemShut {NoStop}%
\bibitem [{\citenamefont {Bulla}\ \emph {et~al.}(1997)\citenamefont {Bulla},
  \citenamefont {Pruschke},\ and\ \citenamefont {Hewson}}]{bulla1997anderson}%
  \BibitemOpen
  \bibfield  {author} {\bibinfo {author} {\bibfnamefont {R.}~\bibnamefont
  {Bulla}}, \bibinfo {author} {\bibfnamefont {T.}~\bibnamefont {Pruschke}}, \
  and\ \bibinfo {author} {\bibfnamefont {A.}~\bibnamefont {Hewson}},\ }\href
  {http://iopscience.iop.org/0953-8984/9/47/014} {\bibfield  {journal}
  {\bibinfo  {journal} {Journal of Physics: Condensed Matter}\ }\textbf
  {\bibinfo {volume} {9}},\ \bibinfo {pages} {10463} (\bibinfo {year}
  {1997})}\BibitemShut {NoStop}%
\bibitem [{\citenamefont {Chen}\ and\ \citenamefont
  {Jayaprakash}(1995)}]{chen1995kondo}%
  \BibitemOpen
  \bibfield  {author} {\bibinfo {author} {\bibfnamefont {K.}~\bibnamefont
  {Chen}}\ and\ \bibinfo {author} {\bibfnamefont {C.}~\bibnamefont
  {Jayaprakash}},\ }\href {http://iopscience.iop.org/0953-8984/7/37/003/}
  {\bibfield  {journal} {\bibinfo  {journal} {Journal of Physics: Condensed
  Matter}\ }\textbf {\bibinfo {volume} {7}},\ \bibinfo {pages} {L491} (\bibinfo
  {year} {1995})}\BibitemShut {NoStop}%
\bibitem [{\citenamefont {Gonzalez-Buxton}\ and\ \citenamefont
  {Ingersent}(1998)}]{gonzalez1998renormalization}%
  \BibitemOpen
  \bibfield  {author} {\bibinfo {author} {\bibfnamefont {C.}~\bibnamefont
  {Gonzalez-Buxton}}\ and\ \bibinfo {author} {\bibfnamefont {K.}~\bibnamefont
  {Ingersent}},\ }\href {\doibase 10.1103/PhysRevB.57.14254} {\bibfield
  {journal} {\bibinfo  {journal} {Phys. Rev. B}\ }\textbf {\bibinfo {volume}
  {57}},\ \bibinfo {pages} {14254} (\bibinfo {year} {1998})}\BibitemShut
  {NoStop}%
\bibitem [{\citenamefont {Vidhyadhiraja}\ \emph {et~al.}(2009)\citenamefont
  {Vidhyadhiraja}, \citenamefont {Macridin}, \citenamefont
  {\ifmmode~\mbox{\c{S}}\else \c{S}\fi{}en}, \citenamefont {Jarrell},\ and\
  \citenamefont {Ma}}]{vidhyadhiraja2009quantum}%
  \BibitemOpen
  \bibfield  {author} {\bibinfo {author} {\bibfnamefont {N.~S.}\ \bibnamefont
  {Vidhyadhiraja}}, \bibinfo {author} {\bibfnamefont {A.}~\bibnamefont
  {Macridin}}, \bibinfo {author} {\bibfnamefont {C.}~\bibnamefont
  {\ifmmode~\mbox{\c{S}}\else \c{S}\fi{}en}}, \bibinfo {author} {\bibfnamefont
  {M.}~\bibnamefont {Jarrell}}, \ and\ \bibinfo {author} {\bibfnamefont
  {M.}~\bibnamefont {Ma}},\ }\href {\doibase 10.1103/PhysRevLett.102.206407}
  {\bibfield  {journal} {\bibinfo  {journal} {Phys. Rev. Lett.}\ }\textbf
  {\bibinfo {volume} {102}},\ \bibinfo {pages} {206407} (\bibinfo {year}
  {2009})}\BibitemShut {NoStop}%
\bibitem [{\citenamefont {Georges}\ \emph {et~al.}(1996)\citenamefont
  {Georges}, \citenamefont {Kotliar}, \citenamefont {Krauth},\ and\
  \citenamefont {Rozenberg}}]{georges1996dynamical}%
  \BibitemOpen
  \bibfield  {author} {\bibinfo {author} {\bibfnamefont {A.}~\bibnamefont
  {Georges}}, \bibinfo {author} {\bibfnamefont {G.}~\bibnamefont {Kotliar}},
  \bibinfo {author} {\bibfnamefont {W.}~\bibnamefont {Krauth}}, \ and\ \bibinfo
  {author} {\bibfnamefont {M.~J.}\ \bibnamefont {Rozenberg}},\ }\href {\doibase
  10.1103/RevModPhys.68.13} {\bibfield  {journal} {\bibinfo  {journal} {Rev.
  Mod. Phys.}\ }\textbf {\bibinfo {volume} {68}},\ \bibinfo {pages} {13}
  (\bibinfo {year} {1996})}\BibitemShut {NoStop}%
\bibitem [{\citenamefont {Gonz\'alez}\ \emph {et~al.}(1999)\citenamefont
  {Gonz\'alez}, \citenamefont {Guinea},\ and\ \citenamefont
  {Vozmediano}}]{gonzalez1999marginal}%
  \BibitemOpen
  \bibfield  {author} {\bibinfo {author} {\bibfnamefont {J.}~\bibnamefont
  {Gonz\'alez}}, \bibinfo {author} {\bibfnamefont {F.}~\bibnamefont {Guinea}},
  \ and\ \bibinfo {author} {\bibfnamefont {M.~A.~H.}\ \bibnamefont
  {Vozmediano}},\ }\href {\doibase 10.1103/PhysRevB.59.R2474} {\bibfield
  {journal} {\bibinfo  {journal} {Phys. Rev. B}\ }\textbf {\bibinfo {volume}
  {59}},\ \bibinfo {pages} {R2474} (\bibinfo {year} {1999})}\BibitemShut
  {NoStop}%
\bibitem [{\citenamefont {Kotov}\ \emph {et~al.}(2012)\citenamefont {Kotov},
  \citenamefont {Uchoa}, \citenamefont {Pereira}, \citenamefont {Guinea},\ and\
  \citenamefont {Castro~Neto}}]{kotov2012electron}%
  \BibitemOpen
  \bibfield  {author} {\bibinfo {author} {\bibfnamefont {V.~N.}\ \bibnamefont
  {Kotov}}, \bibinfo {author} {\bibfnamefont {B.}~\bibnamefont {Uchoa}},
  \bibinfo {author} {\bibfnamefont {V.~M.}\ \bibnamefont {Pereira}}, \bibinfo
  {author} {\bibfnamefont {F.}~\bibnamefont {Guinea}}, \ and\ \bibinfo {author}
  {\bibfnamefont {A.~H.}\ \bibnamefont {Castro~Neto}},\ }\href {\doibase
  10.1103/RevModPhys.84.1067} {\bibfield  {journal} {\bibinfo  {journal} {Rev.
  Mod. Phys.}\ }\textbf {\bibinfo {volume} {84}},\ \bibinfo {pages} {1067}
  (\bibinfo {year} {2012})}\BibitemShut {NoStop}%
\bibitem [{\citenamefont {Jarrell}\ and\ \citenamefont
  {Gubernatis}(1996)}]{jarrell1996bayesian}%
  \BibitemOpen
  \bibfield  {author} {\bibinfo {author} {\bibfnamefont {M.}~\bibnamefont
  {Jarrell}}\ and\ \bibinfo {author} {\bibfnamefont {J.}~\bibnamefont
  {Gubernatis}},\ }\href {\doibase
  http://dx.doi.org/10.1016/0370-1573(95)00074-7} {\bibfield  {journal}
  {\bibinfo  {journal} {Physics Reports}\ }\textbf {\bibinfo {volume} {269}},\
  \bibinfo {pages} {133 } (\bibinfo {year} {1996})}\BibitemShut {NoStop}%
\bibitem [{\citenamefont {Herbut}\ \emph {et~al.}(2009)\citenamefont {Herbut},
  \citenamefont {Juri\ifmmode \check{c}\else \v{c}\fi{}i\ifmmode~\acute{c}\else
  \'{c}\fi{}},\ and\ \citenamefont {Roy}}]{herbut2009theory}%
  \BibitemOpen
  \bibfield  {author} {\bibinfo {author} {\bibfnamefont {I.~F.}\ \bibnamefont
  {Herbut}}, \bibinfo {author} {\bibfnamefont {V.}~\bibnamefont {Juri\ifmmode
  \check{c}\else \v{c}\fi{}i\ifmmode~\acute{c}\else \'{c}\fi{}}}, \ and\
  \bibinfo {author} {\bibfnamefont {B.}~\bibnamefont {Roy}},\ }\href {\doibase
  10.1103/PhysRevB.79.085116} {\bibfield  {journal} {\bibinfo  {journal} {Phys.
  Rev. B}\ }\textbf {\bibinfo {volume} {79}},\ \bibinfo {pages} {085116}
  (\bibinfo {year} {2009})}\BibitemShut {NoStop}%
\bibitem [{\citenamefont {Biroli}\ and\ \citenamefont
  {Kotliar}(2002)}]{biroli2002cluster}%
  \BibitemOpen
  \bibfield  {author} {\bibinfo {author} {\bibfnamefont {G.}~\bibnamefont
  {Biroli}}\ and\ \bibinfo {author} {\bibfnamefont {G.}~\bibnamefont
  {Kotliar}},\ }\href {\doibase 10.1103/PhysRevB.65.155112} {\bibfield
  {journal} {\bibinfo  {journal} {Phys. Rev. B}\ }\textbf {\bibinfo {volume}
  {65}},\ \bibinfo {pages} {155112} (\bibinfo {year} {2002})}\BibitemShut
  {NoStop}%
\bibitem [{\citenamefont {Kyung}\ \emph {et~al.}(2006)\citenamefont {Kyung},
  \citenamefont {Kotliar},\ and\ \citenamefont {Tremblay}}]{Kyung2006}%
  \BibitemOpen
  \bibfield  {author} {\bibinfo {author} {\bibfnamefont {B.}~\bibnamefont
  {Kyung}}, \bibinfo {author} {\bibfnamefont {G.}~\bibnamefont {Kotliar}}, \
  and\ \bibinfo {author} {\bibfnamefont {A.-M.~S.}\ \bibnamefont {Tremblay}},\
  }\href {\doibase 10.1103/PhysRevB.73.205106} {\bibfield  {journal} {\bibinfo
  {journal} {Phys. Rev. B}\ }\textbf {\bibinfo {volume} {73}},\ \bibinfo
  {pages} {205106} (\bibinfo {year} {2006})}\BibitemShut {NoStop}%
\end{thebibliography}%


\end{document}